\documentclass[12pt]{iopart}
\usepackage{epsfig}
\begin{document}

\title[CCM for a family of cylindrical solutions]
{Cauchy-characteristic matching for a family of 
cylindrical solutions possessing both gravitational degrees of freedom}

\author{R\ A\ d'Inverno, M\ R\ Dubal and E\ A\ Sarkies}

\address{Faculty of Mathematical Studies, University of Southampton,
Southampton SO17 1BJ, UK}

\vspace*{0.5pc}
\address{
E-mail: {\tt rdi@maths.soton.ac.uk,}\ {\tt mrd@maths.soton.ac.uk}\ and\ 
{\tt eam@maths.soton.ac.uk} 
}

\begin{abstract}
This paper is part of a long term program to Cauchy-characteristic
matching (CCM) codes as investigative tools in numerical
relativity. The approach has two distinct features: (i) it dispenses
with an outer boundary condition and replaces this with matching
conditions at an interface between the Cauchy and characteristic
regions, and (ii) by employing a compactified coordinate, it proves
possible to generate global solutions. In this paper CCM is applied to
an exact two-parameter family of cylindrically symmetric vacuum solutions
possessing both gravitational degrees of freedom due to Piran, Safier
and Katz. This requires a modification of the previously constructed
CCM cylindrical code because, even after using Geroch decomposition to
factor out the $z$-direction, the family is not asymptotically flat.
The key equations in the characteristic regime turn out to be regular
singular in nature.
\end{abstract}


\section{Introduction}
\label{sec:INTRO}

This paper is part of a long term program to Cauchy-characteristic
matching (CCM) codes as investigative tools in numerical tools in numerical 
relativity. For a review of CCM see the article of Winicour\cite{WIN}. 
The approach has two distinct features: (i) it dispenses with an outer 
boundary condition and replaces this with matching conditions at an 
interface between the Cauchy and characteristic regions, and (ii) by 
employing a compactified coordinate, it proves possible to generate 
global solutions. This paper is continuation of work investigating CCM 
in systems with vacuum cylindrical symmetry (as a prototype system
containing only one spatial degree of freedom). In \cite{CLARKE}
we developed the
necessary machinery and in \cite{DUBAL} we applied the approach to the exact
solution of Weber and Wheeler \cite{WEBER} 
which has one degree of polarization
and so only possesses one gravitational degree of freedom. An agreement
with the exact solution was found with a maximum error better then 1 part
in $10^3$. In the same paper we also investigated the propagation of
Gaussian wave packets possessing two gravitational degrees of freedom, 
but this work did not involve a check against any exact solution. 
Such a test is likely to be a more rigorous one because, unlike the 
Weber-Wheeler case, it would involve passing derivative information 
across the interface.

One of the major problems in numerical relativity is that there are very
few exact solutions known which can be used to test numerical codes. 
In this paper CCM is applied to an exact two-parameter family of 
cylindrically symmetric vacuum solutions possessing both gravitational 
degrees of freedom due to Piran, Safier and Katz \cite{PIRAN}.
This family is somewhat unphysical because the rotational degree of 
freedom diverges at future null infinity. Nonetheless, it can be used to
test the CCM cylindrical code more rigorously because it involves passing 
derivative information across the interface. The previous CCM cylindrical 
code was constructed specifically for solutions which (after using Geroch 
decomposition to factor out the $z$-direction) are asymptotically flat 
and therefore can not be used for processing the Piran et al family. 
However, by working directly in terms of the Geroch potential \cite{GEROCH} 
it proves possible to develop a modified version of the code which is 
able to process the family. In section \ref{sec:TWO} we present the 
family of solutions and briefly discuss its properties. In 
section \ref{sec:CHAR} we review the derivation of the field equations 
in the characteristic region for solutions which, after factoring 
out the $z$-direction, are asymptotically flat. In section
\ref{sec:ASYMP} we discuss the asymptotic limit of the Piran et al 
family and derive the modified field equations and the interface 
equations. An appendix discusses the regular singular nature of two 
of the resulting equations. In section \ref{sec:RESULTS} we briefly 
discuss the results of the modified code. 

\section{The two-parameter family of solutions}
\label{sec:TWO}

Piran et al derive their metric by an indirect method starting from the
Kerr line element in Boyer-Lindquist coordinates 
$(\tilde{t},\tilde{r},\tilde{\theta},\tilde{\phi})$, namely
\begin{equation}
ds^2=-{{\Delta}\over{\rho^2}}(d\tilde{t}
-\tilde{a}\sin^2\tilde{\theta} d\tilde{\phi})^2
+{{\sin^2\tilde{\theta}}\over{\rho^2}}
[(\tilde{r}^2+\tilde{a}^2)d\tilde{\phi}-\tilde{a}d\tilde{t}]^2
+{{\rho^2}\over{\Delta}}d\tilde{r}^2
+\rho^2d\tilde{\theta}^2,
\end{equation}
where
\begin{eqnarray}
\rho^2&=&\tilde{r}^2+\tilde{a}^2\cos^2\tilde{\theta},\\
\Delta&=&\tilde{r}^2-2m\tilde{r}+\tilde{a}^2.
\end{eqnarray}
They first transform to a new isotropic radial coordinate $R$ 
where
\begin{equation}
\tilde{r}=m+R+{{m^2-\tilde{a}^2}\over{4R}},
\end{equation}
and then to cylindrical coordinates 
$(\tilde{t},\tilde{\rho},\tilde{Z},\tilde{\phi})$ 
where
\begin{eqnarray}
\tilde{\rho}&=&R\sin\tilde{\theta},\\
\tilde{Z}&=&R\cos\tilde{\theta}.
\end{eqnarray}
Employing the complex trick of sending
\begin{equation}
\tilde{t}\rightarrow iZ, \quad \tilde{\rho}\rightarrow \tilde{\rho},
\quad \tilde{Z}\rightarrow i\tilde{t},
\quad \tilde{\phi}\rightarrow \tilde{\phi}, \quad \tilde{a}\rightarrow ia,
\end{equation}
they then introduce new cylindrical coordinates $(t,r,z,\phi)$ 
given by
\begin{eqnarray}
t&=&{{\tilde{t}}\over{\alpha}}\Biggl[1
+{{m^2+a^2}\over{4(\tilde{\rho}^2-\tilde{t}^2)}}\Biggr],\\
r&=&{{\tilde{\rho}}\over{\alpha}}\Biggl[1
-{{m^2+a^2}\over{4(\tilde{\rho}^2-\tilde{t}^2)}}\Biggr],\\
z&=&Z-2a^{-1}m(m+\sqrt{m^2+a^2})\tilde{\phi},\\
\phi&=&\alpha\tilde{\phi},
\end{eqnarray}
where
\begin{equation}
\alpha={{\sqrt{m^2+a^2}}\over a},
\end{equation}
and the resulting line element is in the Jordan-Ehlers-Kompaneets
form for a cylindrically symmetric spacetime \cite{JORDON},
\begin{equation}
ds^2=-e^{2\gamma-2\psi}(dt^2-dr^2)+e^{2\psi}(dz+\omega d\phi)^2
+e^{-2\psi}r^2d\phi^2. \label{JEK}
\end{equation}
If we finally introduce coordinates
\begin{eqnarray}
u&=&t-r,\\
v&=&t+r,
\end{eqnarray}
then we can express the two-parameter family of solutions in the form
\begin{equation}
ds^2=-e^{2\gamma-2\psi}dudv+e^{2\psi}(dz+\omega d\phi)^2
+{{e^{-2\psi}(v-u)^2}\over 4}d\phi^2, \label{PSK}
\end{equation}
where explicitly
\begin{eqnarray}
e^{2\gamma}&=&{{(\lambda_u+\lambda_v)^2+\alpha^2(1-\lambda_u \lambda_v)^2}
\over{(1+\lambda_u^2)(1+\lambda_v^2)}}, \label{GAMMA}\\
e^{2\psi}&=&{{\alpha^2(1-\lambda_u\lambda_v)^2+(\lambda_u +\lambda_v)^2}
\over{\alpha^2\Xi^2+(\lambda_u-\lambda_v)^2}}, \label{PSI} \\
\omega&=&{{2a\sqrt{\alpha^2-1}}\over{\alpha}}(\alpha+\sqrt{\alpha^2-1})
\nonumber\\
\phantom{\omega}&\phantom{=}&\phantom{aaaaa}
-{{a\sqrt{\alpha^2-1}\Xi(\lambda_u+\lambda_v)^2}\over
{\sqrt{\lambda_u\lambda_v}[\alpha^2(1-\lambda_u\lambda_v)^2
+(\lambda_u+\lambda_v)^2]}},  \label{OMEGAVAL}
\end{eqnarray}
with
\begin{eqnarray}
\Xi &=& 1+\lambda_u\lambda_v+2\alpha^{-1}\sqrt{\alpha^2-1}
\sqrt{\lambda_u\lambda_v}, \\
\lambda_u&=&\left(\sqrt{a^2+u^2}-u\right)/a, \label{LAMU} \\
\lambda_v&=&\left(\sqrt{a^2+v^2}+v\right)/a. \label{LAMV}
\end{eqnarray}
The coordinates $u$ and $v$ are null, $\phi$ is the canonical azimuthal
coordinate and $z$ lies along the axis of symmetry. 
The two parameters are $a$ $(0\leq a <\infty)$ which is a length scale 
and $\alpha$ $(1\leq\alpha<\infty)$ which is a measure of the total
energy of the system, with $\alpha=1$ corresponding to flat space 
\cite{PIRAN}.
The solution is regular everywhere in the coordinate range
\begin{eqnarray}
-\infty<&u&<\infty,\\
-\infty<&v&<\infty,\\
0\leq&\phi&<2\pi,\\
-\infty<&z&<\infty.
\end{eqnarray}
There is no conical singularity on the axis of symmetry and the solution
reduces to Minkowski spacetime at past and future infinity. However, the
solution is conical at spatial infinity and singular at both past and
future null infinity. More precisely, at future null infinity
(cf Eq.\ (10) in \cite{PIRAN})
\begin{equation}
e^{2\psi} \rightarrow 1, \qquad
e^{2\gamma} \rightarrow {{1+\alpha^2\lambda_u^2}\over{1+\lambda_u^2}},
\qquad \omega \rightarrow -\infty,
\end{equation}
as $v\rightarrow \infty$, with analogous behaviour at past null
infinity on interchanging $u$ and $v$. (Note that Eq.\ (10) in \cite{PIRAN}
includes a typographical error and that $u$ and $v$ should be
transposed). 

Since Piran et al only derived this family of solutions
by the indirect method 
described above, we tried to confirm that the family 
is vacuum by a direct computation of the Ricci tensor
using both of the computer algebra systems SHEEP \cite{SHEEP} and 
GRTENSOR \cite{GRTENSOR}.
Unfortunately, neither system was able to complete the 
calculation because the metric involves nested radicals and algebra
systems are notorious for the difficulties such quantities present. 
However, we were able to confirm the indirect derivation described
above. 

\section{The standard field equations}
\label{sec:CHAR}

The standard treatment of the Jordan-Ehlers-Kompaneets cylindrically
symmetric line element (\ref{JEK}) in the Cauchy region
\begin{eqnarray}
t_0&\leq& t\leq t_f, \\
0&\leq& r\leq 1,
\end{eqnarray}
is described in \cite{DUBAL}. 
In particular, it is shown that the independent set of dynamical
equations for the variables $\psi$, $\omega$, $L^\phi_z$ and 
$\tilde{L}$ are
\begin{eqnarray}
\psi_{,t} &=& \frac{1}{r} {\tilde L},
\label{PSIDOTL} \\
\omega_{,t} &=&  -2e^{-4\psi} L^\phi_z,
\label{OMEGADOTL} \\
L^\phi_{z,t} &=&  \frac{1}{r} e^{4\psi} (\frac{1}{2} \omega_{,r} - 
\frac{1}{2}
r \omega_{,rr} -2r\psi_{,r} \omega_{,r}),
\label{LPHIZDOT} \\
{\tilde L}_{,t} &=&  \frac{1}{r} [r^2\psi_{,rr} + r\psi_{,r} - 
\frac{1}{2}
e^{4\psi} \omega_{,r}^2 + 2 e^{-4\psi} (L^\phi_z)^2] ,
\label{LTILDEDOT}
\end{eqnarray}
where $L^\phi_z$ and $\tilde{L}$ are defined in terms of the mixed
components of the extrinsic curvature $K^\mu_\nu$ by
\begin{eqnarray}
L^\phi_z&=&r^2e^{\gamma-\psi}K^\phi_z, \\
\tilde{L}&=&r^2e^{\gamma-\psi}(K^\phi_\phi-\omega K^\phi_z).
\end{eqnarray}
These equations are augmented by the constraint equation
\begin{equation}
\gamma_{,r} = \frac{1}{4r} e^{4\psi}\omega_{,r}^2 - 
r\psi_{,r}^2+\frac{1}{r}[\tilde{L}^2+e^{-4\psi}(L^\phi_z)^2],
\label{HAML}
\end{equation}
which serves to determine $\gamma$ once the main variables are known.

\begin{figure}
\epsfig{file=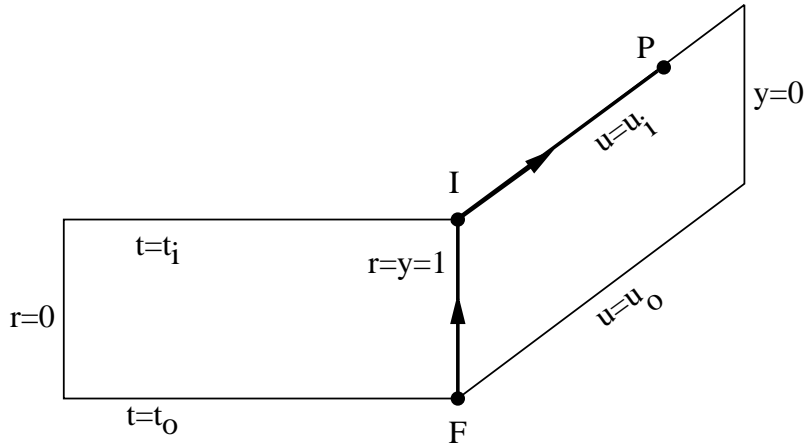}
\caption{Evaluation of the Geroch potential, $o$, at a general point $P$
requires integration along the path $F\rightarrow I \rightarrow P$, see
text for details.}
\label{FIG1}
\end{figure}

In the rest of this section we review the treatment in the
characteristic region since this will need modification to cope with the
Piran et al family of solutions.  We introduce the compactified 
coordinate
\begin{equation}
y={1\over {\sqrt{r}}}={{\sqrt{2}}\over{\sqrt{v-u}}},
\end{equation}
in which case the line element becomes
\begin{equation}
ds^2=-e^{2(\gamma-\psi)}du^2
+{4\over y^3}e^{2(\gamma-\psi)}dudy
+e^{2\psi}(dz+\omega d\phi)^2
+{e^{-2\psi}\over {y^4}}d\phi^2, \label{NJEK}
\end{equation}
where the metric functions $\psi$, $\omega$ and $\gamma$ are now to be
regarded as functions of $u$ and $y$. The characteristic region then
consists of
\begin{eqnarray}
u_0&\leq&u\leq u_f, \\
0&\leq&y\leq 1,
\end{eqnarray}
where the interface is taken to be at $r=y=1$ and future null infinity
is given by $y=0$. Instead of working directly with the metric functions
$\psi$ and $\omega$, we use the related quantities $m$ and $w$ which are
defined by
\begin{eqnarray}
m&=&{{e^{2\psi}-1}\over y},\\
w&=&{o\over y}=
-\int_F^P\lambda^2(\textstyle{1\over 2}y^4\omega_{,y}+y\omega_{,u})du
+\displaystyle{{1\over y}\int_F^P\lambda^2y^2\omega_{,y}dy},
\label{OMCHARDEF}
\end{eqnarray}
where
\begin{equation}
\lambda=e^{2\psi}=1+my,
\end{equation}
$o$ is the Geroch potential \cite{GEROCH}, and the integration is along any
path connecting a fixed point $F$ on the interface (we choose
$(u,y)=(u_0,1)$) to a general point $P$ in the characteristic region
(see Fig.\ \ref{FIG1}).
With these definitions the vacuum equations can be written in the
succinct form
\begin{eqnarray}
M&=&m_{,u}/\lambda,\label{E1}\\
W&=&w_{,u}/\lambda,\label{E2}\\
M_{,y}&=&f(y,m,m_{,y},m_{,yy},w,w_{,y},W),\label{E3}\\
W_{,y}&=&g(y,m,m_{,y},w,w_{,y},w_{,yy},M),\label{E4}
\end{eqnarray}
where $f$ and $g$ are explicit functions of their arguments 
\cite{DUBAL}.
Eqs.\ (\ref{E1}) and (\ref{E2}) serve to define $M$ and $W$ and 
(\ref{E3}) and (\ref{E4}) provide coupled propagation equations for $M$
and $W$ along the null rays ruling the hypersurfaces 
$u=\mbox{constant}$. The initial data consists of specifying $m$ and $w$
on the initial hypersurface $\{u=u_0, \ 0\leq y\leq 1\}$ and the interface
$\{u_0\leq u\leq u_f,\ y=1\}$, together with $\gamma$ at their intersection
$(u,y)=(u_0,1)$. The essence of the iterative integration scheme is to use 
(\ref{E3}) and (\ref{E4}) to determine $M$ and $W$ on 
the interior null hypersurface $u=u_i$
and then (\ref{E1}) and (\ref{E2}) to determine $m_{,u}$ and $w_{,u}$
on $u=u_i$, which in turn determines $m$ and $w$ on the next
neighbouring null slice. Finally $\gamma$ is determined from the
constraint equation (Eq.\ (42) in \cite{DUBAL}) which has the form
\begin{equation}
\gamma_{,y}=h(y,m,m_{,y},w,w_{,y}),
\label{E5}
\end{equation}
where $h$ is an explicit function of its arguments. The CCM method 
requires the exchange of values of the metric functions and their 
derivatives at the interface at each iteration, the details of which are
given in \cite{DUBAL}.

\section{The modified field equations}
\label{sec:ASYMP}

The approach described in the last section breaks down for the Piran 
et al solution in the characteristic region because 
$\omega \rightarrow -\infty$ as $y \rightarrow 
0$. Let us consider the behaviour of the various functions on the
hypersurface $u=u_i$ in the asymptotic limit $y \rightarrow 0$.
We have,
\begin{eqnarray}
\lambda_u&=&(\sqrt{a^2+u_i^2}-u_i)/a,\\
\lambda_v&=&\bar{\lambda}_v/ y^2,\\
\Xi&=&\bar{\Xi}/ y^2,
\end{eqnarray}
where we set
\begin{eqnarray}
\bar{\lambda}_v&\equiv&\left(2+u_iy^2
+\sqrt{4+4u_iy^2+(a^2+u_i^2)y^4}\right)/a =4/a+O(y^2),\\
\bar{\Xi}&\equiv&\lambda_u\bar{\lambda}_v
+2y\sqrt{(1-\alpha^{-2})\lambda_u\bar{\lambda}_v}+y^2
=4\lambda_u / a+O(y).
\end{eqnarray}
Thus, asymptotically,
\begin{eqnarray}
e^{2\psi}&=&1+O(y),\\
e^{2\gamma}&=&(1+\alpha^2\lambda_u^2)/(1+\lambda_u^2)+O(y^2),\\
\omega&=&-\Biggl({{2\sqrt{(\alpha^2-1)a\lambda_u}}\over{1
+\alpha^2\lambda_u^2}}\Biggr){1\over y}
+{{2a\sqrt{\alpha^2-1}}\over{\alpha}}\left(\alpha+\sqrt{\alpha^2-1}\right)
+O(y), \label{OMEGDIV}
\end{eqnarray}
and
\begin{eqnarray}
m&=&{{2\alpha\sqrt{(\alpha^{2}-1)a\lambda_u^3}}\over {1+\alpha^2\lambda_u^2}}
+O(y),\label{Em}\\
w&=&{b\over y}+{{2\sqrt{(\alpha^2-1)a\lambda_u}}\over
{1+\alpha^2\lambda_u^2}}+O(y),\\
o&=&b+{{2\sqrt{(\alpha^2-1)a\lambda_u}}\over
{1+\alpha^2\lambda_u^2}}y+O(y^2),\label{Eo}
\end{eqnarray}
where
\begin{equation}
b=o(u_i,0).
\end{equation}
It is clear that $\lambda_v$, $\Xi$, $\omega$ and $w$ are all divergent
as $y\rightarrow 0$. However, the ancillary quantities 
$\bar{\lambda}_v$ and $\bar{\Xi}$ as well as $m$ and $o$,
the Geroch potential, are all regular. This suggests rewriting the 
system (\ref{E1})--(\ref{E5}) in terms of these last two variables.
We find explicitly that
\begin{eqnarray}
M&=&{{m_{,u}}\over{1+my}},\label{E6}\\
O&=&{{o_{,u}}\over{1+my}},\label{E7}\\
M_{,y}&=&-{{o_{,y}}\over{y(1+my)}}O
+{1\over{4(1+my)}}\Biggl[-y(m+y^2m_{,yy}+3ym_{,y})
\nonumber\\
\phantom{M_y}&\phantom{=}&\phantom{aaaaaaaaaaaaa}
+{{y^2}\over{1+my}}(m^2+2ymm_{,y}
+y^2m_{,y}^2-o_{,y}^2)\Biggr]\label{E8},\\
O_{,y}&=&{1\over y}O+{{yo_{,y}}\over{(1+my)}}M
-{{y^2}\over{4(1+my)}}(yo_{,yy}+o_{,y})
\nonumber\\
\phantom{M_y}&\phantom{=}&\phantom{aaaaaaaaaaaaaaa}
+{{y^3}\over{2(1+my)^2}}(mo_{,y}+ym_{,y}o_{,y}),
\phantom{aaaa}
\label{E9}\\
\gamma_{,y}&=&-{y\over{8(1+my)^2}}\Bigl((m+ym_{,y})^2+o_{,y}^2\Bigr).
\label{E10}
\end{eqnarray}
Although the equations (\ref{E8}) and (\ref{E9}) are coupled singular
equations for $M$ and $O$, they are regular singular equations and 
the solutions remain regular as $y\rightarrow 0$ (see the Appendix). 
In fact, it is clear from the defining equations (\ref{E6}) and
(\ref{E7}), together with (\ref{Em}) and (\ref{Eo}), that both $M$ 
and $O$ are of order unity in the limit $y\rightarrow 0$. These 
are the modified equations on which the new code is based for 
investigating the Piran et al solutions.

We need to augment the equations with the interface equations which
relate the two sets of variables in the two coordinate systems on the
interface. The relationships required for {\it injection}, i.e. going
from the characteristic to the Cauchy region are,
\begin{eqnarray}
\psi&=&{1\over 2}\ln(1+my) ,
\label{PSIMREL} \\
\psi_{,r}&=&-{1\over 2}yM -{1\over4}{{y^3}\over{\lambda}}(ym)_{,y}, \\
\psi_{,rr}&=&{1\over 2}yM_{,u}+{1\over 2}y^3(yM)_{,y}
 -{1\over 8}{{y^6}\over{\lambda^2}} [(ym)_{,y}]^2 
\nonumber\\
\phantom{M_y}&\phantom{=}&\phantom{aaaaaaaaaaaaaaaaa}
 +{1\over 8}{{y^5}\over{\lambda}} 
 \left\{2(ym)_{,y}+ [y(ym)_{,y}]_{,y}\right\}, \\
\tilde{L}&=&{1\over 2}{M\over y},   \\
\omega_{,r}&=&{O\over{y^2\lambda}}, \\
\omega_{,rr}&=&-{{O_{,u}}\over{y^2\lambda}}+{{MO}\over{y\lambda}}
+{{O}\over{\lambda}}
+{1\over 2}{{yO(ym)_{,y}}\over {\lambda^2}}
-{1\over 2}{{yO_{,y}}\over \lambda}, \\
L^{\phi}_{z}&=&{1\over 2}{{\lambda O}\over{y^2}}+{1\over 4}yo_{,y}.
\end{eqnarray}
Similarly, the relationships required for {\it extraction}, i.e. going
from the Cauchy to the characteristic region are,
\begin{eqnarray}
\lambda&=&e^{2\psi} , \\
m&=&r^{1/2}(e^{2\psi}-1) , 
\label{MPSIREL} \\
m_{,y}&=&-r^2e^{2\psi}(4\psi_{,t}+4\psi_{,r}+r^{-1})+r , \\
m_{,yy}&=&2r^{7/2}e^{2\psi}(4\psi_{,tt}+8\psi_{,tr}+8\psi^2_{,t}
+10r^{-1}\psi_{,t}+16\psi_{,t}\psi_{,r}+4\psi_{,rr}  
\nonumber\\
\phantom{M_y}&\phantom{=}&\phantom{aaaaaaaaaaaaaaaaaaa}
+8\psi^2_{,r}+10r^{-1}\psi_{,r}+r^{-2}) -2r^{3/2},  
\label{MYY} \\
M&=&2r^{-1/2}\tilde{L}, 
\label{LTMREL} \\
o&=&\int_F^I r^{-1}e^{4\psi}\omega_{,r}dt, 
\label{pot} \\
o_{,y}&=&4r^{1/2}L^{\phi}_z-2r^{1/2}e^{4\psi}\omega_{,r}, \\
o_{,yy}&=&
2r^2e^{4\psi}(8\psi_{,t}\omega_{,r}+2\omega_{,tr}
+r^{-1}\omega_{,r}+8\psi_{,r}\omega_{,r} 
+2\omega_{,rr}+r^{-1}\omega_{,t})
\nonumber\\
\phantom{M_y}&\phantom{=}&\phantom{aaaaaaaaaaaaaaaaaaaaaaaaaaaaaaaa}
-8r^2(L^{\phi}_{z,t}+L^{\phi}_{z,r}), \\
O&=&r^{-1}e^{2\psi}\omega_{,r} .
\label{LWREL}
\end{eqnarray}
where the integration in (\ref{pot}) is along the interface from the
point $F: (t,r)=(t_0,1)$ to the point $I: (t,r)=(t_i,1)$ as shown in Fig.\ 
\ref{FIG1}.

\section{Results}
\label{sec:RESULTS}

In order to test the CCM code against the Piran et al solution we will
first compare the metric quantity $\psi$ in both the 
Cauchy and characteristic regions. Eq.\ (\ref{PSI}) provides the 
exact solution in both cases when $\lambda_u$ and $\lambda_v$, Eqs.\ 
(\ref{LAMU}) and (\ref{LAMV}), are written in terms of $(t,r)$ coordinates
and $(u,y)$ coordinates for the Cauchy and characteristic regions 
respectively. 
It is not possible to use $\omega$ for a similar comparison since
from (\ref{OMEGDIV}) we have seen that it diverges as $y\rightarrow 0$. 
We use, instead, the Geroch potential, $o$, which
is computed directly by the code in the 
characteristic region, but must be constructed in the Cauchy region using 
its definition (cf Eq.\ (24) of \cite{CLARKE}),
\begin{equation}
o = \int r^{-1} e^{4\psi}\omega_{,r}dt + \int r^{-1}e^{4\psi}
\omega_{,t}dr. \label{GERCAUCHY}
\end{equation}
It is straightforward to evaluate the integrals in the Cauchy region
using finite-difference representations of $\omega_{,r}$ and $\omega_{,t}$.
For the exact solution $\omega_{,r}$ and $\omega_{,t}$ can be evaluated
by differentiation, however the integral must be evaluated
numerically to find a semi-analytic $o$. A similar approach is 
required for the `exact' value of $o$ in the characteristic region 
using Eq.\ (\ref{OMCHARDEF}).

As mentioned previously, the two parameters $a$ and $\alpha$ of the Piran
et al solution represent the length scale and strength of the gravitational
wave respectively. For $t < 0$ the wave moves in from infinity, reaches
its highest concentration at $t=0$, where it rebounds from the $r=0$ axis,
and is outgoing for $t>0$. CCM allows us to evolve initial data containing
an ingoing gravitational wave, something which is difficult to do for a
Cauchy only evolution. Since we have set the interface at $r=1$ the value
of $a$ can be chosen such that a substantial fraction of the 
the wave will move onto the Cauchy region and will be well resolved with the 
numbers of grid points we choose. Changing the interface position while 
keeping the `interface distance to $a$' ratio the same results in an 
identical evolution on the Cauchy portion of the grid.

The primary variable we used for comparison are  $\psi$ and $o$ (since
$\gamma$ is a derived quantity). We considered grid resolution 
numbers of $N=301, 601$ and $1201$, where the error is defined as
\begin{equation}
\epsilon(\psi) = ||\psi_E-\psi_C||_2/||\psi_E||_2,
\label{NORMDEF}
\end{equation}
where $\psi_E$ is the exact value, $\psi_C$ is the code computed value and 
$||\cdot||_2$ denotes the $L_2$ norm. We ran the modified code for a whole
range of values in the $a, \alpha$ parameter space and found an error in 
$\psi$ of no more than 0.01\% and an error in $o$ of less than 0.2\%.
However, in this version of the code, although the convergence rate of 
the solution starts as second-order it deteriorates to first order after
long time evolution. In some recent work, colleagues in the Southampton
Numerical Relativity group have modified the code so that it is now
fully second order and very long time stable \cite{SPER}.
The improvement was achieved  by dispensing with
variables which are related by exponential or logarithmic functions 
at the interface, using the Geroch potential both in the Cauchy 
and characteristic regimes and also using an implicit method at the 
interface. This version of the code is currently being applied successfully 
to modelling dynamic cosmic strings.

\begin{figure}
\epsfig{file=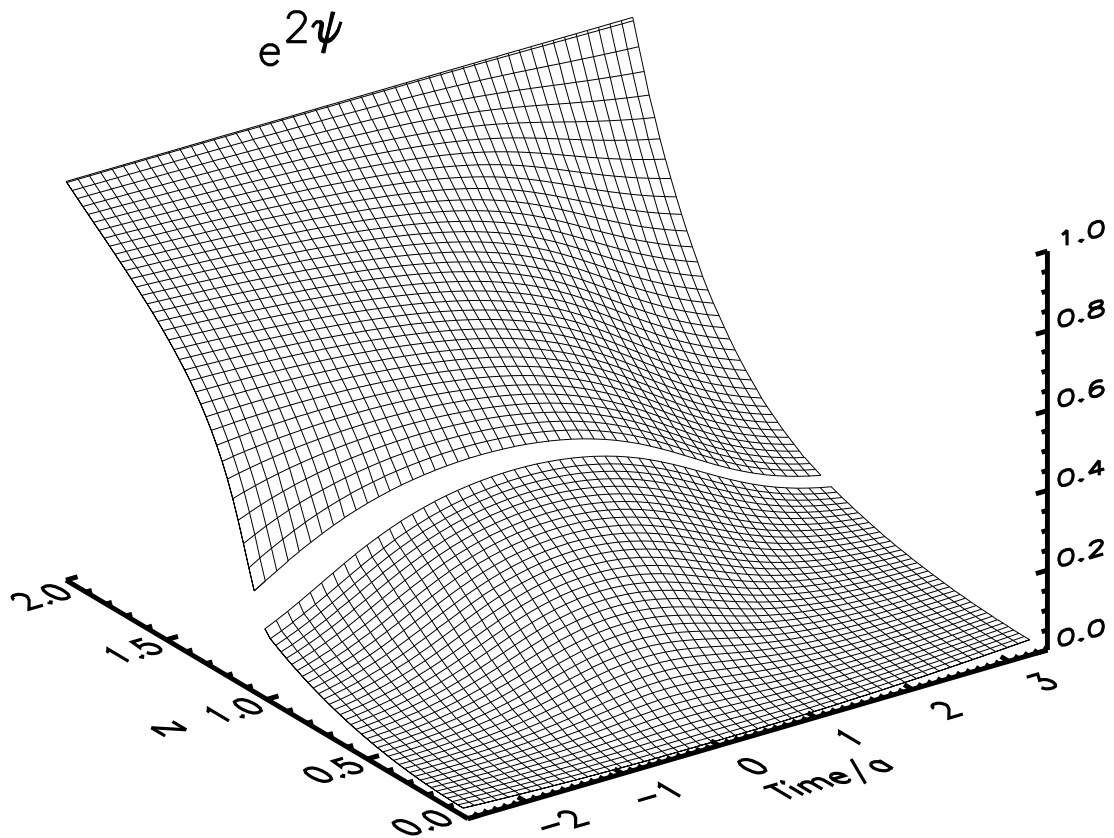}
\caption{Surface plot showing the time evolutions of the metric 
quantity $e^{2\psi}$. The initial data has the parameters $a=0.5$
and $\alpha=10.0$, representing a strong field spacetime. The interface
is placed at $z=1$ denoted by a gap in the surface plot. Note that for
clarity the $z$-axis is reversed in this figure.}
\label{FIG2a}
\end{figure}

\begin{figure}
\epsfig{file=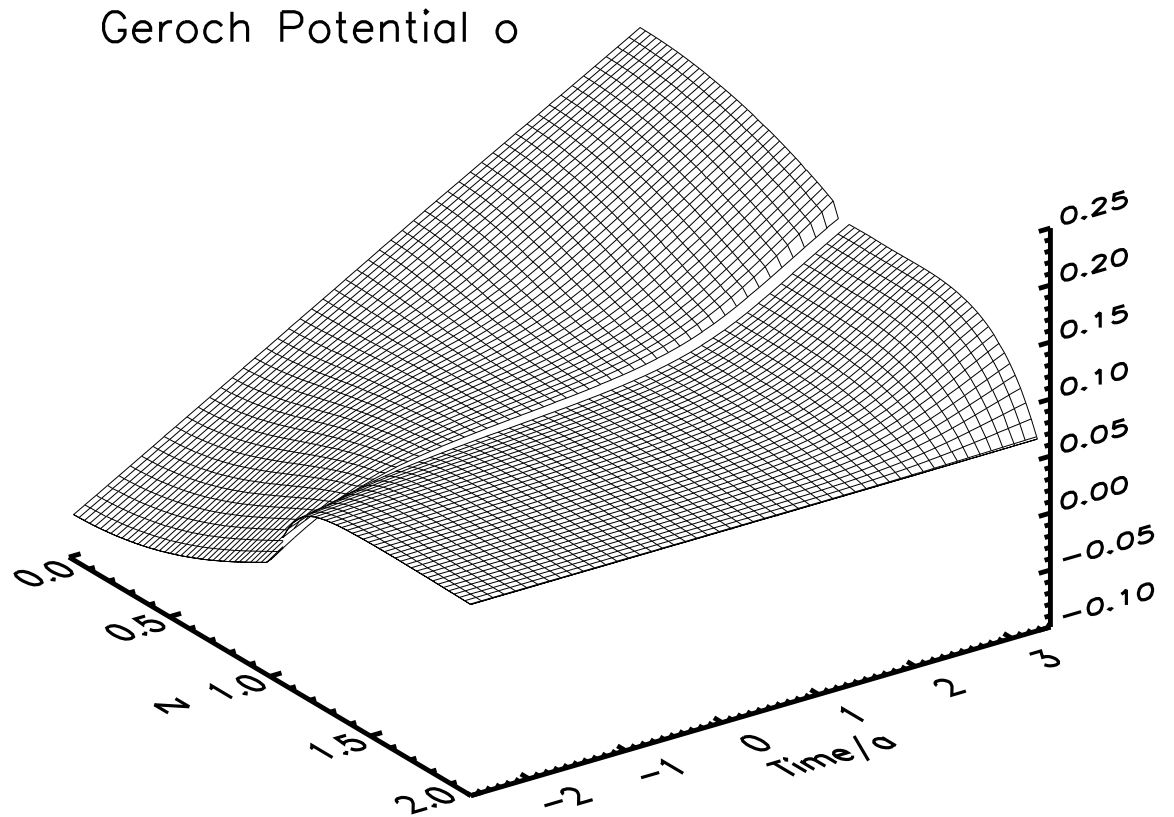}
\caption{Surface plot showing the time evolutions of the 
Geroch potential, $o$.}
\label{FIG2b}
\end{figure}

\begin{figure}
\epsfig{file=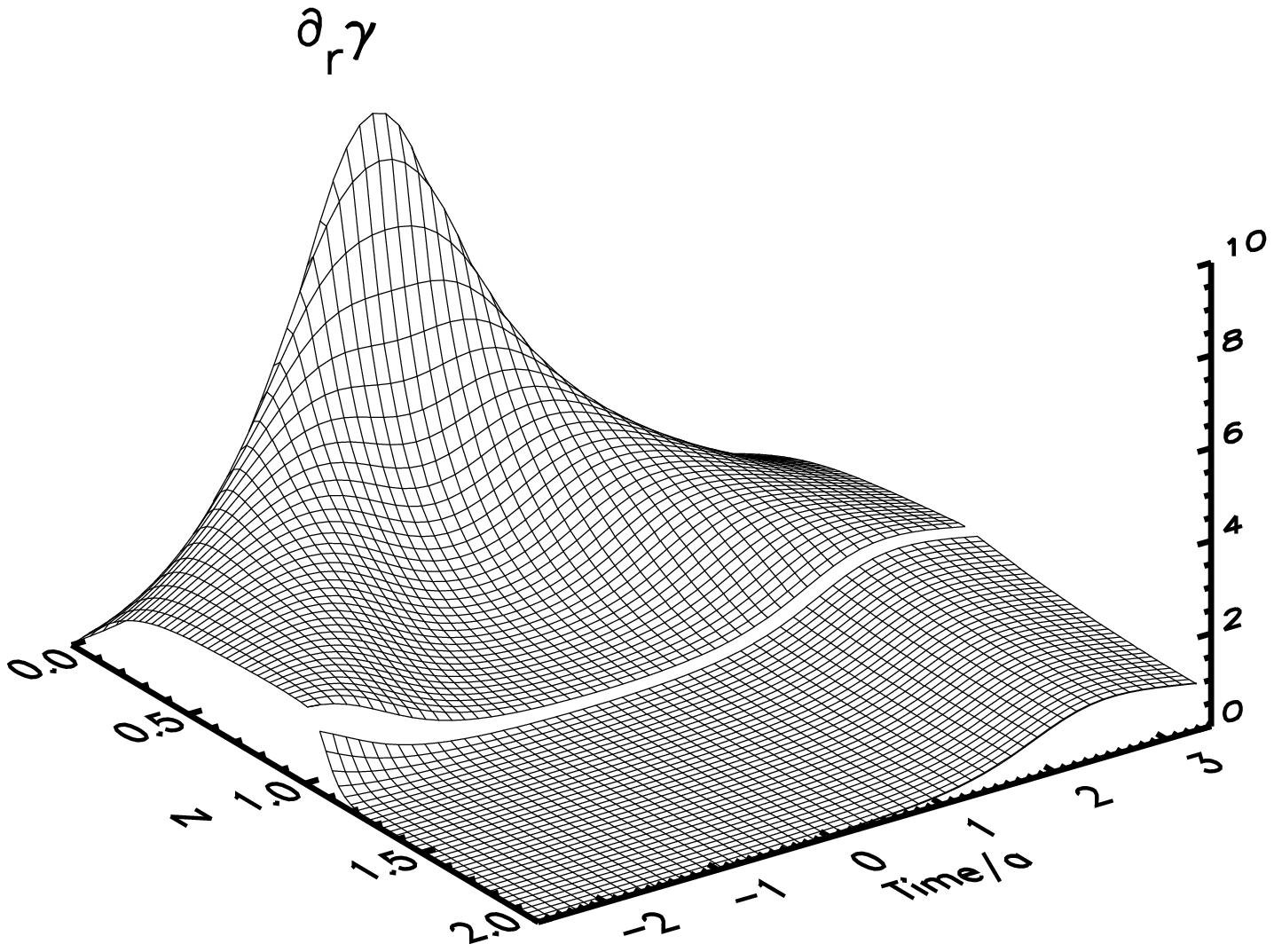}
\caption{Surface plot showing the time evolutions of the energy 
distribution, $\gamma_{,r}$. }
\label{FIG2c}
\end{figure}

Surface plots are shown for evolutions of 
the metric quantity $e^{2\psi}$ in Fig.\ \ref{FIG2a}, 
the Geroch potential $o$ in Fig.\ \ref{FIG2b},
and the radial derivative $\gamma_{,r}$ (which indicates the
distribution of energy within the wave)in Fig.\ \ref{FIG2c}, 
where the parameters are $a=0.5$
and $\alpha=10$. The surface plots use a radial coordinate $z$ defined by,
\begin{equation}
z = \left\{\begin{array}{ll}
     r & \mbox{for $0 \le r \le 1$} \\
     2-y & \mbox{for $0 \le y \le 1$}
     \end{array} \right.
\end{equation}
so that there is a change in the coordinate system at $r=y=z=1$.
Note that the value of $e^{2\psi}$ remains equal to 1 at null
infinity, but is very close to zero at the axis $r=0$. The incoming wave
hits the axis at $t=0$ and rebounds. This can be seen very clearly in
the surface plot of $\gamma_{,r}$. The large peak of Fig.\ 2c grows 
without bound as $\alpha$ is increased. 

\begin{figure}
\epsfig{file=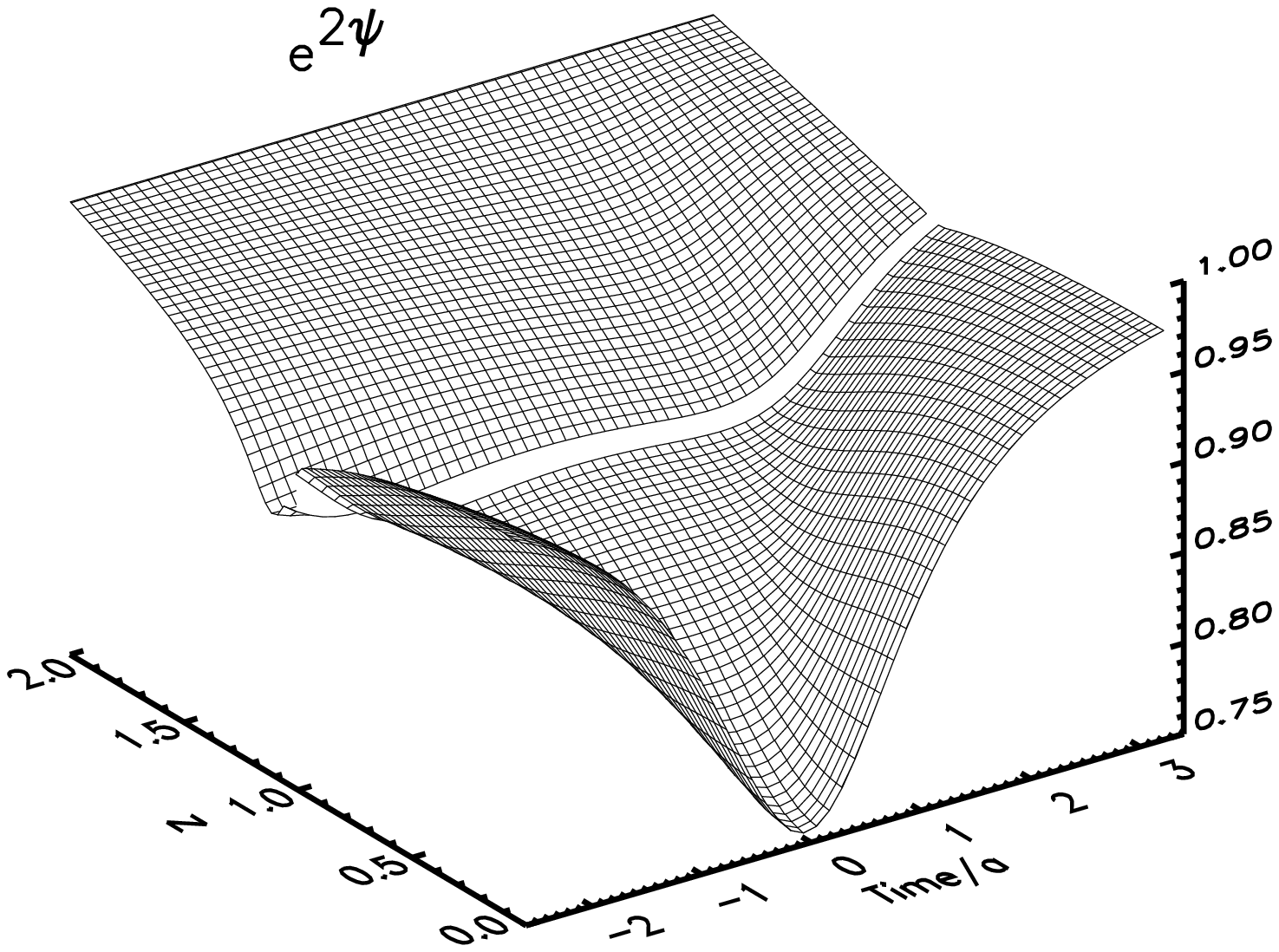}
\caption{Same as Fig.\ \ref{FIG2a}, but with parameters $a=0.5$ and 
$\alpha=1.01$ representing an almost flat spacetime.}
\label{FIG3a}
\end{figure}

\begin{figure}
\epsfig{file=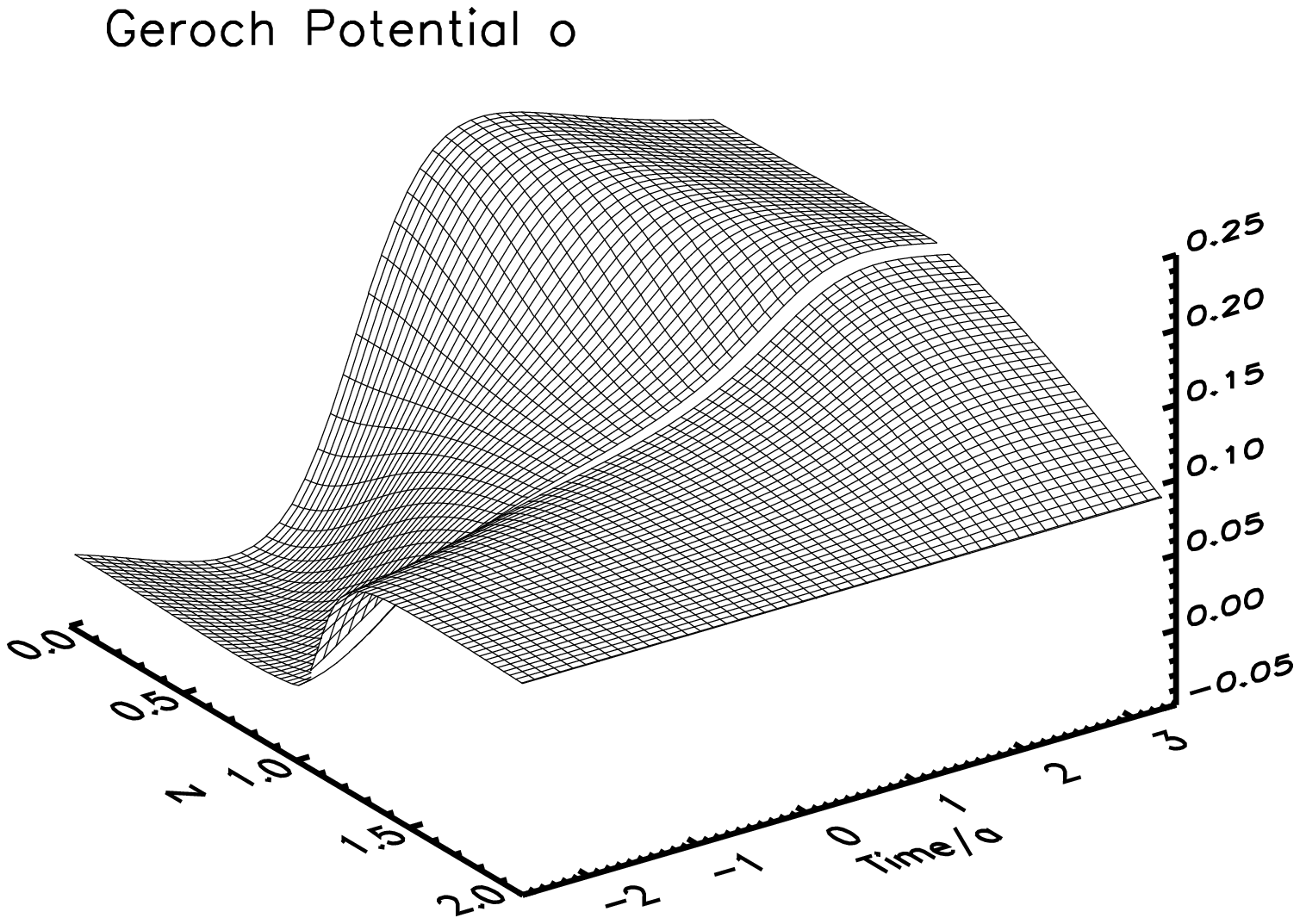}
\caption{Same as Fig.\ \ref{FIG2b}, but with parameters $a=0.5$ and 
$\alpha=1.01$ representing an almost flat spacetime.}
\label{FIG3b}
\end{figure}

\begin{figure}
\epsfig{file=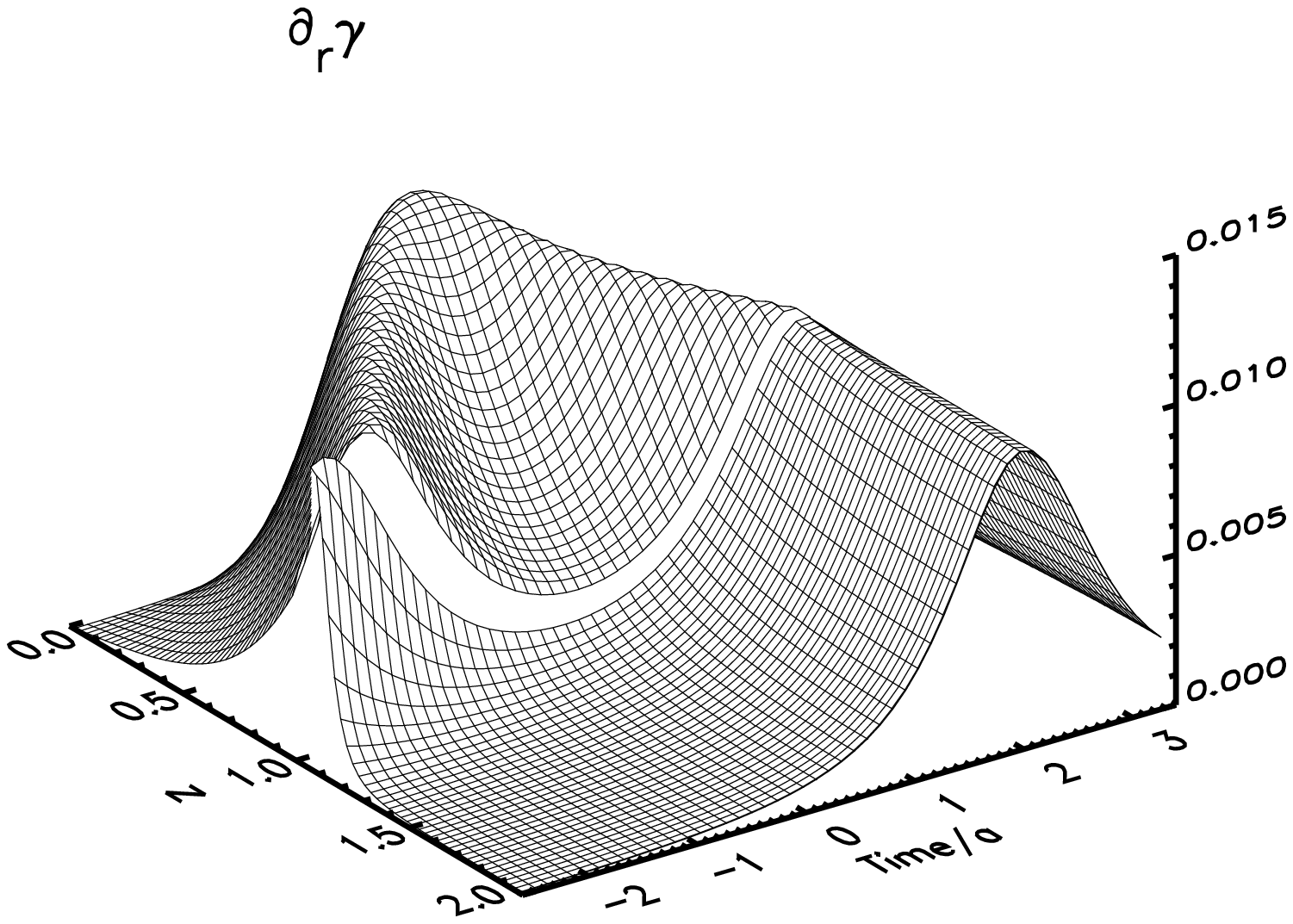}
\caption{Same as Fig.\ \ref{FIG2c}, but with parameters $a=0.5$ and 
$\alpha=1.01$ representing an almost flat spacetime.}
\label{FIG3c}
\end{figure}

For comparison purposes figures \ \ref{FIG3a}, \ \ref{FIG3b} and
\ \ref{FIG3c} shows the same 
quantities, but for parameter values $a=0.5$ and $\alpha=1.01$, 
which represents an almost flat spacetime. Again the wave reaches its
maximum concentration on the $r=0$ axis at $t=0$, however the peak is 
much less dominant in this case.

\ack{We would like to thank James Vickers for helpful discussions.
This work has been supported by PPARC grant number GR/K44510.}

\appendix
\section{Regular-singular behaviour of modified propagation equations}

Let
\begin{equation}
F={{o_{,y}}\over{(1+my)}},
\end{equation}
which by (\ref{Em}) and (\ref{Eo}) is regular as $y\rightarrow 0$.
The homogeneous part of the modified propagation equations 
(\ref{E8}) and (\ref{E9}) is
\begin{eqnarray}
M_{,y}+{F\over y}O&=&0,\label{A1}\\
O_{,y}-yFM-{1\over y}O&=&0.\label{A2}
\end {eqnarray}
Differentiating (\ref{A2}) with respect to $y$ and using (\ref{A2})
to eliminate $M$ and (\ref{A1}) to eliminate $M_{,y}$ we get
\begin{equation}
O_{,yy}-\Biggl({2\over y}+{{F_{,y}}\over F}\Biggr)O_{,y}
+\Biggl({2\over {y^2}}+{{F_{,y}}\over{yF}}+F^2\Biggr)O=0,
\end{equation}
which has dominant singular part
\begin{equation}
O_{,yy}-{2\over y}O_{,y}+{2\over {y^2}}O=0.
\end{equation}
Substituting in the trial solution $y^k$ we obtain the auxiliary
equation
\begin{equation}
k^2-3k+2=0
\end{equation}
which has roots $k=1,2$ and so gives rise to the regular independent
solutions $y$ and $y^2$.

\end{document}